\documentclass[sigconf,screen]{acmart}

\AtBeginDocument{%
  \providecommand\BibTeX{{%
    \normalfont B\kern-0.5em{\scshape i\kern-0.25em b}\kern-0.8em\TeX}}}

\copyrightyear{2021}
\acmYear{2021}
\setcopyright{acmcopyright}\acmConference[BuildSys '21]{The 8th ACM International Conference on Systems for Energy-Efficient Buildings, Cities, and Transportation}{November 17--18, 2021}{Coimbra, Portugal}
\acmBooktitle{The 8th ACM International Conference on Systems for Energy-Efficient Buildings, Cities, and Transportation (BuildSys '21), November 17--18, 2021, Coimbra, Portugal}
\acmPrice{15.00}
\acmDOI{10.1145/3486611.3486648}
\acmISBN{978-1-4503-9114-6/21/11}

\usepackage{lineno}
\modulolinenumbers[5]
\usepackage{url}
\usepackage{amsmath}

\usepackage[ruled,vlined]{algorithm2e}
\usepackage{subcaption}
\usepackage{multirow}
\usepackage{comment}

\usepackage{xcolor}

\begin{document}
\title{Improving Load Forecast in Energy Markets During COVID-19}


\author{Ziyun Wang}
\affiliation{%
  \department{Faculty of Science}
  \institution{National University of Singapore}
  \city{}
  \state{}
  \country{Singapore}
  \postcode{Zipcode if any}
}
\email{wang.ziyun@u.nus.edu}

\author{Hao Wang}
\authornote{Corresponding author: \url{hao.wang2@monash.edu} (Hao Wang).}
\affiliation{%
  \department{Department of Data Science and AI}
  \department{and Monash Energy Institute}
  \institution{Monash University}
  \city{Melbourne}
  \state{Victoria}
  \country{Australia}
  \postcode{3800}
}
\email{hao.wang2@monash.edu}


\begin{abstract}
The abrupt outbreak of the COVID-19 pandemic was the most significant event in 2020, which had profound and lasting impacts across the world. Studies on energy markets observed a decline in energy demand and changes in energy consumption behaviors during COVID-19. However, as an essential part of system operation, how the load forecasting performs amid COVID-19 is not well understood. This paper aims to bridge the research gap by systematically evaluating models and features that can be used to improve the load forecasting performance amid COVID-19. Using real-world data from the New York Independent System Operator, our analysis employs three deep learning models and adopts both novel COVID-related features as well as classical weather-related features. We also propose simulating the stay-at-home situation with pre-stay-at-home weekend data and demonstrate its effectiveness in improving load forecasting accuracy during COVID-19.
\end{abstract}




\maketitle

\section{Introduction}
At the end of 2019, a novel infectious coronavirus (named COVID-19) was identified, which quickly developed into a worldwide pandemic in 2020 \cite{zhou_2020}.
Many governments issued stay-at-home (or shelter-in-Place) orders to contain the spread of the coronavirus, asking people to self-quarantine until they had the disease under control \cite{moreland_2020,xu_2020}. The sudden changes in daily life had an impact on the electricity industry around the world, and many countries have observed significant changes in electricity consumption patterns \cite{iea_2021}. Researchers in \cite{aburayash_2020,santiago_2021,eryilma_2020} have attempted to record and review the impact of COVID-19 on the electricity industry, but there was a research gap on load forecasting during COVID-19. It is not well understood how the load forecast performs during an emergency event like the COVID-19 pandemic and how to improve the performance of load forecasting. \emph{We are thus motivated to bridge the research gap by systematically evaluating models and features that can be used to improve load forecasting amid COVID-19.}

During the pandemic, decreasing electricity consumption was observed worldwide \cite{iea_2021}. Amid COVID-19, a study in Ontario, Canada revealed that the overall electricity demand has declined by 14\% in April 2020 \cite{aburayash_2020}, Spain displayed a decreased power demand of 13.49\% due to the lockdown measures \cite{santiago_2021}, while in the US, overall electricity generation declined after the issue of stay-at-home advisories \cite{eryilma_2020}. 
The limit on mobility adds further complexity to electricity consumption behaviors. Under the full lockdown scenario, increased residential consumption and decreased industrial electricity consumption were observed in a lot of places in the world, such as Lagos, Nigeria \cite{edomah_2020}, Kragujevac, Serbia \cite{cvetkovic_2021}, China \cite{cheshmehzangi_2020}, and New York, US \cite{chen_2020}. The study on the residential electricity consumption in Kragujevac, Serbia also identified an increase in consumption for heating and electrical devices brought about by the increased presence of residents in their households \cite{cvetkovic_2021}. Despite having such volatile load demand profiles during the pandemic, only a few studies were conducted on improving load forecasting accuracy. Among them was a study by Chen et al. \cite{chen_yang_2020}, according to which mobility data forms a useful feature for load forecasting during COVID-19, which will be examined in this paper as well.

Short-term load forecasting is an essential tool for operating reserve \cite{santos_2013}, which is an extra generating power prepared for any form of supply disruption. Since the operating reserve is at least the capacity of the largest supplier plus a fraction of the peak load \cite{wang_2005_operating}, underestimation of peak load can be disastrous, while overestimation will cause utility companies to waste extra resources in larger operating reserve \cite{kim_2019_short}. Load forecasting accuracy is crucial for utility companies to avoid further loss with the fickle energy consumption during the pandemic.

This paper aims to systematically analyze load forecasting models and relevant features that can improve load forecasting in an emergency event like COVID-19. 
We take the energy market of New York Independent System Operator (NYISO) in the US as our case study, as the New York state was at the epicenter of the pandemic during the first wave in 2020 \cite{washington_2020}. Since we target the wholesale market instead of regional systems, the mix of electric and gas heating is not changing significantly and is not explicitly considered in our model.
The main contributions of this paper are the follows.
\begin{itemize}
    \item This paper is among the first to systematically evaluates the effectiveness of COVID-19 related features and deep learning models for load forecasting during COVID-19.
    \item We propose using pre-stay-at-home weekend data as extra training data, which effectively simulates post-stay-at-home situations and improves load forecasting accuracy when COVID-19 hits.
    \item We compare Fully Connected DNN (FCDNN), Long Short-Term Memory (LSTM), and Gated Recurrent Unit (GRU) models, displaying comparable performance under different forecasting scenarios during COVID-19.
    \item We adopt COVID-19 cases and mobility data as novel auxiliary features for load forecasting during COVID-19 and validate their effectiveness.
\end{itemize}
Our work suggests a significant improvement of load forecasting methods in an emergency like the pandemic. Insights such as simulating stay-at-home situations with weekend data and adopting mobility features for load forecasting during COVID-19 can help stakeholders, such as system operators and utilities, better prepare for similar contingencies in the future.

\section{Data Description}
The data used for load forecasting in this paper is imported from COVID-EMDA+ data hub \cite{ruan_2020} and COVID-19 Community Mobility Reports (CCMR) \cite{google_2021}, and we introduce the data in detail as follows. 

\subsection{Load Data}
\label{load_data}
Load data $L_t$ is at the core of load forecasting, where $t$ indicates time step. We extract load data from the COVID-EMDA+ data hub, which contains hourly load profiles measured in MW for the NYISO energy market. Weather and COVID case data in Section \ref{feature_data} are also from the same data hub and will be introduced later.

\begin{figure}[!b]
    \centering
    \vspace{-2mm}
    \includegraphics[width=0.95\columnwidth]{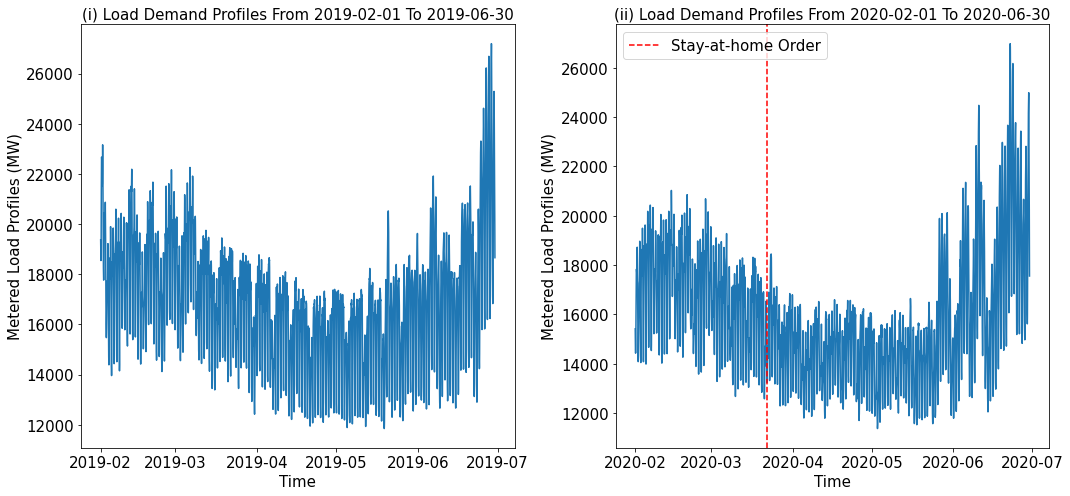}
    \caption{Visualization of the NYISO energy demand.}
    \label{fig:load}
\end{figure}

The stay-at-home order on March 22nd, 2020 in New York state \cite{moreland_2020} splits $L_t$ into pre-stay-at-home $L^{pre}_t$ and post-stay-at-home $L^{post}_t$. Figure \ref{fig:load} shows lower energy demand in NYISO after the stay-at-home order compared to the same period in 2019. Since the load data dates back to the earliest in 2017, we use data from Jan 1st, 2017 to May 31st, 2020. The major impact of stay-at-home order causing people not reporting to work resembles the situation of weekends. Hence, we also propose using pre-stay-at-home weekend data $L^{weekend}_t$ to simulate the pattern of $L^{post}_t$ in Section \ref{weekend}.

\subsection{Weather, COVID Cases and Mobility Data}
\label{feature_data}
Traditional load forecasting utilized weather as auxiliary information. We acquire hourly weather data $W_t$, including air temperature, dew point temperature, wind speed, and relative humidity. To mimic the effect of day-ahead weather forecasts, we shift the weather data backward by 24 hours.

We evaluate the effectiveness of COVID case data $C_t$ and mobility data $M_t$ in assisting load forecast. $C_t$ includes daily new confirmed and death cases, signaling the severity of the pandemic. $M_t$ contains daily workplace and residential mobility, indicating the impact of quarantine and work-from-home policies. $M_t$ is obtained from \cite{google_2021}, measuring percentage change in visitors to workplaces or time spent in residential areas with respect to the median value from a 5-week period, from Jan 3rd to Feb 6th, 2020.

While $C_t$ and $M_t$ are daily records, $L_t$ and $W_t$ are measured hourly. For consistency in format, each data entry in $C_t$ and $M_t$ is repeated 24 times to simulate hourly measurements. Moreover, the first non-zero data entry of $C_t$ and $M_t$ is in Jan and Feb 2020, respectively, later than that of $L_t$ and $W_t$. Therefore, we fill the empty data entries in $C_t$ and $M_t$ with zeros so that their sizes match that of $L_t$ and $W_t$.

\section{Methodology Description}
Load forecasting can employ statistical models for time series analysis, or deep learning models that are promising in tackling sequential data \cite{hosein_2017}. We apply both statistical models, such as Auto-Regressive Integrated Moving Average (ARIMA) and Deep Neural Networks, including FCDNN, LSTM, and GRU, to load forecasting and evaluate their performance. We also evaluate the performance of pre-stay-at-home weekend training data $L^{weekend}_t$ as suggested in Section \ref{load_data} against that of ordinary training data $L^{pre}_t$.

\subsection{Benchmark Scenario}
\label{benchmark}
We need to build a benchmark model by mimicking the scenario, in which no changes are made to improve load forecasting accuracy in the face of COVID-19. We train the aforementioned models with a classical feature set, including load $L^{pre}_t$ and weather $W^{pre}_t$, to produce day-ahead load forecasts for $L^{post}_t$ for $10$ weeks. The forecast results are then evaluated against actual values using the Mean Absolute Percentage Error (MAPE),
\begin{equation}
    \text{MAPE} = \dfrac{100}{n} \sum^n_{t=1} \left\lvert\dfrac{L_t-F_t}{L_t}\right\rvert
\end{equation}
where $L_t$ and $F_t$ are actual and forecast data points at time step $t$, respectively, and $n$ is the number of fitted points.

\begin{figure}[!t]
    \centering
    \includegraphics[width=0.85\columnwidth]{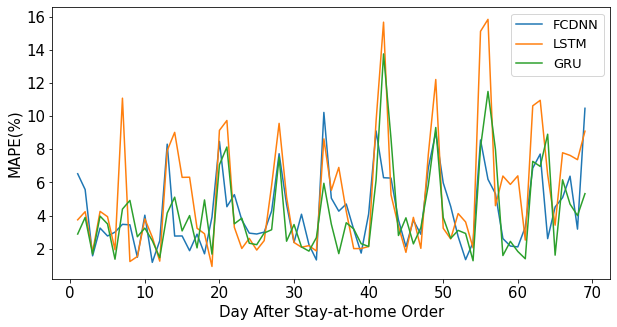}
    \caption{Preliminary performance of FCDNN, LSTM, and GRU in the benchmark scenario.}
    \label{fig:prelim}
\end{figure}

FCDNN, LSTM, and GRU exhibit comparable performance as demonstrated in Figure \ref{fig:prelim}, achieving an overall MAPE of 4.48\%, 5.32\%, and 4.85\% for the $10$-week period after the stay-at-home order, respectively. Although the overall accuracy of the three models seems to remain in the acceptable range, they are much worse than their performance under the ordinary pre-stay-at-home situation, in which they achieved 2.97\%, 3.22\%, and 2.78\%, respectively. In Figure \ref{fig:prelim}, the daily forecasting MAPEs also fluctuate drastically for all three models, reaching high peaks $>10\%$ around day $40$ to $60$, suggesting volatile performance forecasting unusual $L^{post}_t$ data. 

Among all the evaluated models, ARIMA is unable to produce meaningful results as the resultant accuracy is much lower than that of the other three models. Thus we will not consider ARIMA in the actual analysis. Since LSTM has been demonstrated to be effective in load forecasting per \cite{hossain_2020, muzaffar_2019, wang_2021}, we choose the prevalent LSTM model as our benchmark result for later comparisons.

\subsection{Simulating Stay-at-home Impact Using Weekend Load Data}
\label{weekend}
As mention in Section \ref{load_data}, we propose to use $L^{weekend}_t$ to capture the impact of the Stay-at-home policy. We believe that $L^{weekend}_t$ improves the efficiency of training by removing the influence of weekday data which are more dissimilar to $L^{post}_t$. To verify this hypothesis, we trained FCDNN, LSTM, and GRU with $L^{weekend}_t$ and $W^{weekend}_t$ and used them to perform day-ahead load forecasts for $L^{post}_t$ for $10$ weeks, and the resultant MAPEs are compared against our benchmark result in Section \ref{benchmark}.

\begin{figure}[htbp]
    \centering
    \includegraphics[width=0.85\columnwidth]{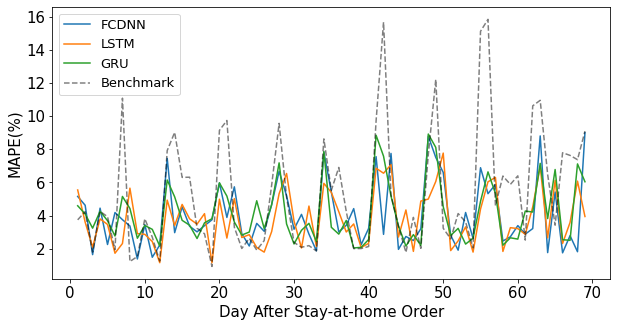}
    \caption{Load forecast performance using pre-stay-at-home weekend load data.}
    \vspace{-4mm}
    \label{fig:weekend}
\end{figure}

As shown in Figure \ref{fig:weekend} and Table \ref{tab:weekend}, FCDNN, LSTM, and GRU models trained with $L^{weekend}_t$ and $W^{weekend}_t$ showed comparable performance with a significant improvement in accuracy from the benchmark results, achieving an overall MAPE of 4.04\%, 3.85\%, and 4.13\% respectively forecasting $10$ weeks of $L^{post}_t$. This shows that simulating $L^{post}_t$ with $L^{weekend}_t$ is an effective measure in improving load forecasting accuracy during COVID-19. Therefore, in the following sections, $L^{weekend}_t$ will be used instead of $L^{pre}_t$ as training data.

\begin{table}[htbp]
    \centering
    \caption{Load forecasting overall accuracy using pre-stay-at-home weekend data.}
    \begin{tabular}{ c  c c c c }
    \hline
    Scenarios & Benchmark & FCDNN & LSTM & GRU \\
    \hline
    MAPE (\%) & 5.32 & 4.04 & 3.85 & 4.13 \\
    \hline 
    \end{tabular}
    \label{tab:weekend}
\end{table}

\section{Case Study}
Using the NYISO energy market as a case study, we aim to systematically compare the effectiveness of different feature sets, $W_t$, $C_t$ and $M_t$, and different models, including FCDNN, LSTM, and GRU, in load forecasting during COVID-19. 

\subsection{COVID-19 Related Auxiliary Features}
\label{features}
We propose using COVID-related features such as COVID cases $C_t$ and mobility $M_t$ as indicators of notable changes during the pandemic, and compare their effectiveness against $W_t$ as auxiliary features for the day-ahead forecasts of $L^{post}_t$.

Since $C_t$ and $M_t$ are blank for the pre-stay-at-home period, we have to provide more information of $C^{post}_t$ and $M^{post}_t$ in order to train the models. Therefore, we employ a rolling forecast technique, such that $L^{post}_\tau$ is added to the training data $L^{weekend}_t$ as stated in Section \ref{weekend}, where $\tau$ denotes the week index after the stay-at-home order and $L^{post}_\tau = \{L^{post}_t \mid 7(\tau-1) \leq t \leq 7\tau\}$, to produce day-ahead load forecasts for $L^{post}_{\tau+1}$. Both $C_t$ and $M_t$ experience drastic changes immediately after stay-at-home order, so day-ahead forecasts for $L^{post}_{\tau=1}$ using $C_t$ and $M_t$ will reasonably display high inaccuracies. This gives $C_t$ and $M_t$ a natural disadvantage, which is their disability to be used immediately after stay-at-home order.

\begin{figure*}[!t]
\centering
    \begin{subfigure}{0.33\textwidth}
        \centering
        \includegraphics[width=\textwidth]{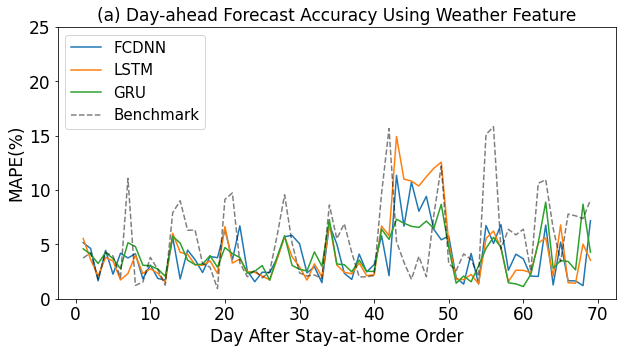}
    \end{subfigure}
    \begin{subfigure}{0.33\textwidth}
        \centering
        \includegraphics[width=\textwidth]{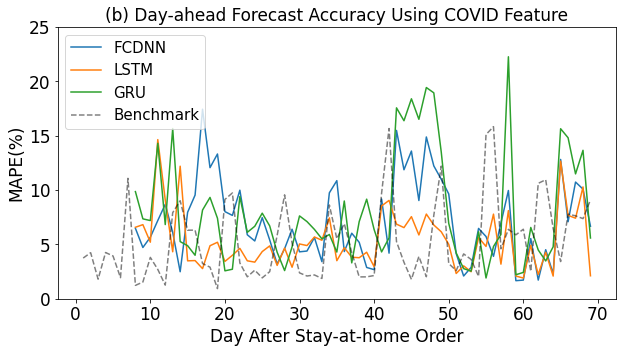}
    \end{subfigure}
    \begin{subfigure}{0.33\textwidth}
        \centering
        \includegraphics[width=\textwidth]{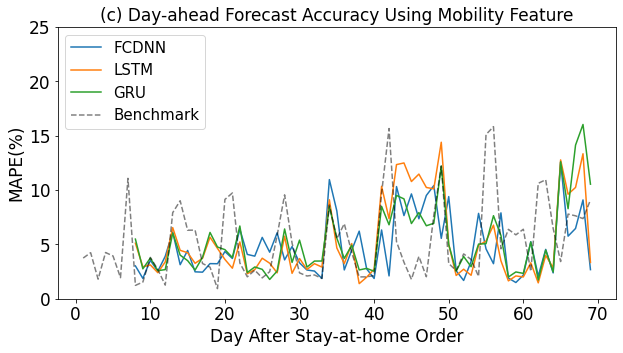}
    \end{subfigure}
\caption{Rolling forecast using weather, COVID cases and mobility auxiliary features.}
\label{fig:features}
\end{figure*}

As shown in Figure \ref{fig:features}, the rolling forecast technique is vulnerable to distortion during day $40$ to $50$ (in Week 7) after the stay-at-home order, across all auxiliary features used. Compared to Figure \ref{fig:weekend}, rolling forecast trained with the same feature $W_t$ performs comparably well for the undistorted days, but the additional information from $L^{post}_{\tau=6}$ hindered the judgment of the models, causing soaring forecasting MAPEs.

Moving on, $M_t$ as an auxiliary feature has displayed a certain level of effectiveness in load forecasting during COVID-19. Compared to benchmark results, $M_t$ shows significant improvement in load forecasting accuracy except for $L^{post}_{\tau=7}$ and $L^{post}_{\tau=10}$ and has comparable performance to that of $W$ during the same period. This coincides with the insights from \cite{chen_yang_2020} that mobility data can be an effective feature in load forecasting during COVID-19. On the other hand, $C_t$ is rather baleful than helpful in load forecasting for $L^{post}_t$, showing a higher MAPE than benchmark result across almost the entire $10$-week period. Even if the COVID case data is capable of indicating latent changes during the pandemic, it is less directly related to load demand profiles. It hence is less effective than $W_t$ and $M_t$ when used in load forecasting.

\begin{table}[htbp]
    \centering
    \caption{Overall MAPEs of forecasts using weather, COVID cases and mobility auxiliary features.}
    \begin{tabular}{c c c c}
    \hline
    Features & Weather & COVID & Mobility \\
    \hline
    FCDNN & 4.08\% & 7.15\% & 4.84\% \\
    LSTM & 4.26\% & 5.38\% & 5.21\% \\
    GRU & 4.04\% & 8.05\% & 5.26\% \\
    \hline
    Benchmark & 5.32\% & - & - \\
    \hline
    \end{tabular}
    \label{tab:features}
\end{table}

\section{Conclusion and Future Work}
Using the NYISO market as a case study, we evaluated models and features that can be used to improve the accuracy of load forecasting amid COVID-19. We provided insights into improving load forecast in future contingencies similar to COVID-19 as follows.
\begin{itemize}
    \item We used the pre-stay-at-home weekend data to simulate the stay-at-home situation, and showed that it effectively improved load forecasting accuracy compared to using ordinary pre-stay-at-home training data. 
    \item We then adopted auxiliary features, including COVID cases and mobility data, and compared their performance with the classical weather feature. In comparison, the mobility feature showed improvements in forecast accuracy though slightly, the COVID feature was clearly not as helpful. 
    \item Across all scenarios analyzed as stated above, we evaluated the performance of deep learning models, including FCDNN, LSTM, and GRU, which showed comparative performance.
\end{itemize}

Our key finding is that pre-stay-at-home weekend load data led to significant improvement in load forecast during COVID-19 and can be easily obtained compared to COVID cases and mobility data.

In our future work, we plan to explore data from more energy markets with distinct energy consumption behaviors and evaluate the various combinations of different feature sets. We are also interested in exploring potential causes for the distorted rolling forecasts as shown in Section \ref{features}.

\section{ACKNOWLEDGMENTS}
This work is in part supported by the FIT Academic Staff Funding of Monash University. The authors thank anonymous reviewers for their helpful comments and suggestions and Dr. Yize Chen for informative discussions.

\bibliographystyle{ACM-Reference-Format}
\bibliography{ref.bib}

\end{document}